\documentclass[runningheads]{llncs}
\usepackage{graphicx}
\usepackage{epsfig}
\usepackage{subfigure}
\usepackage{xcolor}
\begin{document}
\title{Deep Interactive Learning: An Efficient Labeling Approach for Deep Learning-Based Osteosarcoma Treatment Response Assessment}
\titlerunning{Deep Interactive Learning}
%
\author{David Joon Ho\inst{1,3} \and
Narasimhan P. Agaram\inst{1,3} \and
Peter J. Sch\"uffler\inst{1} \and
Chad M. Vanderbilt\inst{1} \and
Marc-Henri Jean\inst{1} \and
Meera R. Hameed\inst{1,4} \and
Thomas J. Fuchs\inst{1,2,4}}

\authorrunning{D. J. Ho et al.}
%
\institute{Department of Pathology, Memorial Sloan Kettering Cancer Center, New York, NY 10065 USA\\
\email{hod@mskcc.org}\and
Weill Cornell Graduate School for Medical Sciences, New York, NY 10065 USA\and
The first two authors contributed equally\and
The last two authors contributed equally
}



\maketitle              
\begin{abstract}
Osteosarcoma is the most common malignant primary bone tumor. 
Standard treatment includes pre-operative chemotherapy followed by surgical resection.
The response to treatment as measured by ratio of necrotic tumor area to overall tumor area is a known prognostic factor for overall survival. 
This assessment is currently done manually by pathologists by looking at glass slides under the microscope which may not be reproducible due to its subjective nature. 
Convolutional neural networks (CNNs) can be used for automated segmentation of viable and necrotic tumor on osteosarcoma whole slide images. 
One bottleneck for supervised learning is that large amounts of accurate annotations are required for training which is a time-consuming and expensive process. 
In this paper, we describe Deep Interactive Learning (DIaL) as an efficient labeling approach for training CNNs. 
After an initial labeling step is done, annotators only need to correct mislabeled regions from previous segmentation predictions to improve the CNN model until the satisfactory predictions are achieved. 
Our experiments show that our CNN model trained by only 7 hours of annotation using DIaL can successfully estimate ratios of necrosis within expected inter-observer variation rate for non-standardized manual surgical pathology task. 
\end{abstract}
\section{Introduction}
Osteosarcoma is the most common bone cancer occurring in adolescents with a second smaller peak in older adults \cite{ottaviani2009}. 
Pre-operative chemotherapy followed by surgery is a standard treatment for osteosarcoma. 
The ratio of necrotic tumor to overall tumor post neoadjuvant chemotherapy is a well-known prognostic factor and correlates with patients' survival \cite{huvos1977,rosen1982}. 
Thus, for patients with localized disease who have undergone complete resection, if the ratio of tumor necrosis is greater than 90\%, the 5-year survival is higher than 80\%. 
Currently, the ratio of tumor necrosis is manually estimated by pathologists by microscopic review of multiple glass slides from resected specimens.

Computational pathology has provided automated and reproducible techniques to analyze digitized histopathology images \cite{fuchs2011}, especially with convolutional neural networks (CNNs) \cite{srinidhi2019}.
Arunachalam showed a patch-level classification CNN composed of three  convolutional layers and two fully-connected layers could be used to identify viable tumor, necrotic tumor, and non-tumor in osteosarcoma \cite{arunachalam2019}.
For more accurate analysis, fully convolutional networks were developed for a pixel-wise classification, also known as semantic segmentation \cite{long2015}.
U-Net segmenting subcellular structures in microscopy images was described in \cite{ronneberger2015}.
More recently, Deep Multi-Magnification Network (DMMN) was introduced for multi-class tissue segmentation of histopathology images by looking at patches in multiple magnifications and has shown outstanding segmentation performance in breast cancer \cite{ho2019}.

Performance of these supervised machine learning methods highly depends on the amount and quality of annotations.
Public datasets with annotations are generally available for common cancer types such as breast \cite{ehteshami2017,bandi2019} and have been widely used for training CNNs \cite{wang2016,lee2018}.
For rare cancers such as osteosarcoma, fresh manual annotations by pathologists with specialized expertise are required.  Such annotations require a lot of time from busy professionals and thus optimizing for reduced burden is paramount.
To reduce annotation time, interactive learning has been developed.
Interactive learning allows annotators to ``interact'' with a machine learning model by correcting predictions of the model to improve its performance until the predictions are satisfied \cite{fails2003,schueffler2013}.
An interactive segmentation toolkit for biomedical images, known as ilastik, was introduced in \cite{sommer2011,berg2019}.
Here, random forest classifiers \cite{breiman2001} were used for segmentation.
QuPath \cite{bankhead2017} was developed to interactively analyze giga-pixel whole slide images where segmentation was also done based on random forest classifiers \cite{breiman2001}.

In this paper, we propose Deep Interactive Learning (DIaL) by integrating the concept of interactive learning into deep learning framework for multi-class tissue segmentation of histopathology images and treatment response assessment for osteosarcoma.
To evaluate our segmentation model, we estimate the necrosis ratio in case-level by counting the number of pixels predicted as viable tumor and necrotic tumor by the segmentation model and compare with the ratio from pathology reports.
We observe our CNN model can estimate the necrosis ratio within expected inter-observer variation rate for non-standardized manual surgical pathology task. 
Note that the total labeling time took approximately 7 hours with DIaL.

\section{Proposed Method}
\begin{figure}[t!]
\centerline{\epsfig{figure=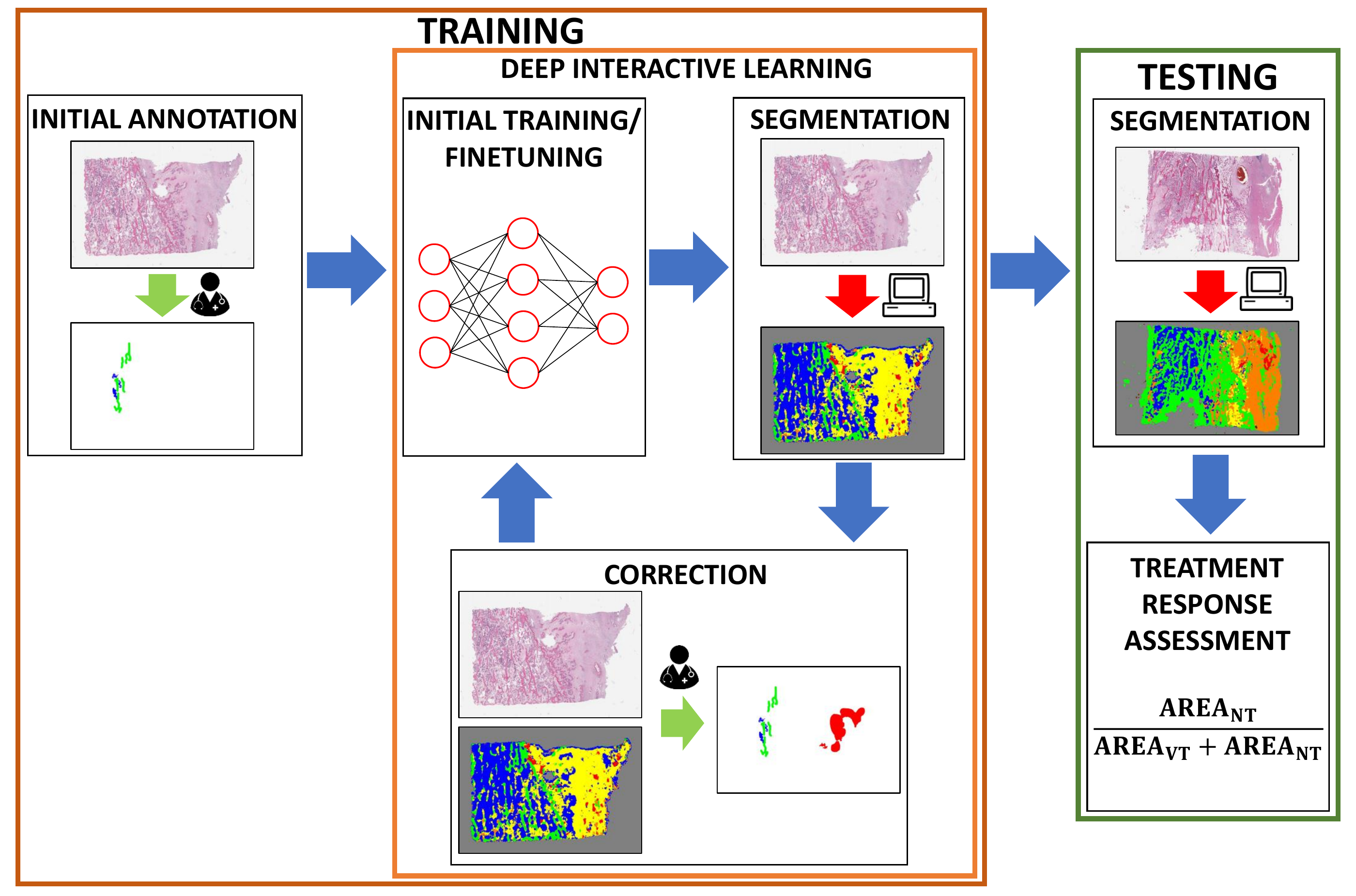,width=\textwidth}}
\caption{Block diagram of the proposed method. First of all, initial annotation is done on training whole slide images (WSIs) where characteristic features of each class are partially annotated. The annotated regions are used to train a Deep Multi-Magnification Network \cite{ho2019}. Segmentation is done on the same training WSIs to correct any mislabeled regions containing challenging or rare features. These corrected regions are added to the training set to finetune the model. This training-segmentation-correction iteration, denoted as Deep Interactive Learning (DIaL), is repeated until segmentation predictions are satisfied by annotators. The final model is used to segment testing WSIs to assess treatment responses.}
\label{fig:block}
\end{figure}
It is necessary to manually label osteosarcoma whole slide images (WSIs) to supervise a segmentation convolutional neural network (CNN) for automated treatment response assessment.
Labeling WSIs exhaustively would be ideal but it needs tremendous labeling time.
Partial labeling approaches are introduced to reduce labeling time \cite{bokhorst2019,ho2019}, but challenging or rare morphological features can be missed.
We propose Deep Interactive Learning (DIaL) to efficiently annotate both characteristic features and challenging features on WSIs to have outstanding segmentation performance.
Our block diagram is shown in Figure \ref{fig:block}.
First of all, initial annotation is partially done mainly on characteristic features of classes.
During DIaL, training a CNN, segmentation prediction, and correction on mislabeled regions are repeated to improve segmentation performance until segmentation predictions on training images are satisfied by the annotators.
Note that challenging or rare features would be labeled during the correction step.
When training the CNN is finalized, the CNN is used to segment viable tumor and necrotic tumor on testing cases to assess treatment responses.

\subsection{Initial Annotation}
\label{sec:initial_annotation}
\begin{figure}[ht]
\centering
\subfigure[]{\frame{\epsfig{figure=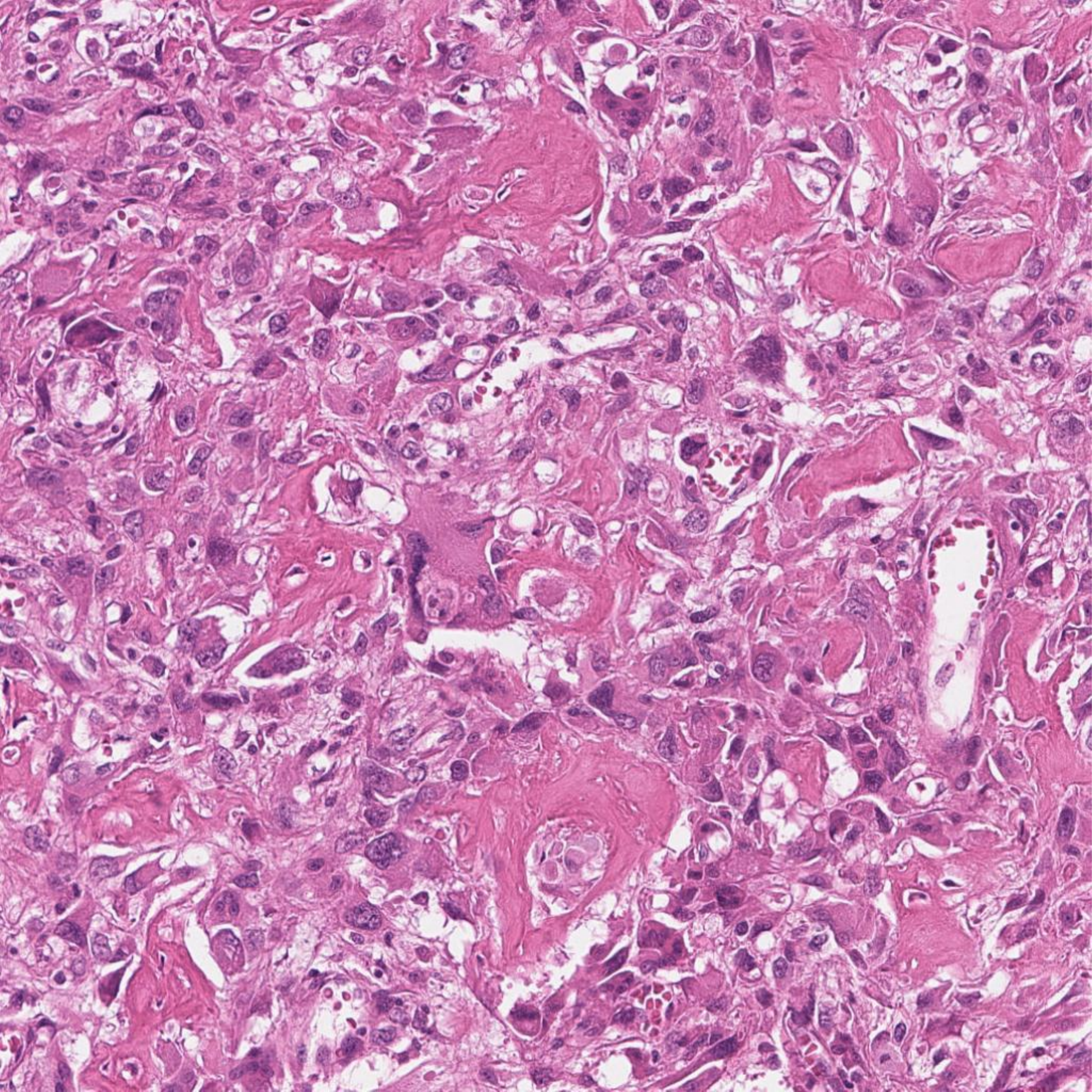,width = 0.13\textwidth}}}
\subfigure[]{\frame{\epsfig{figure=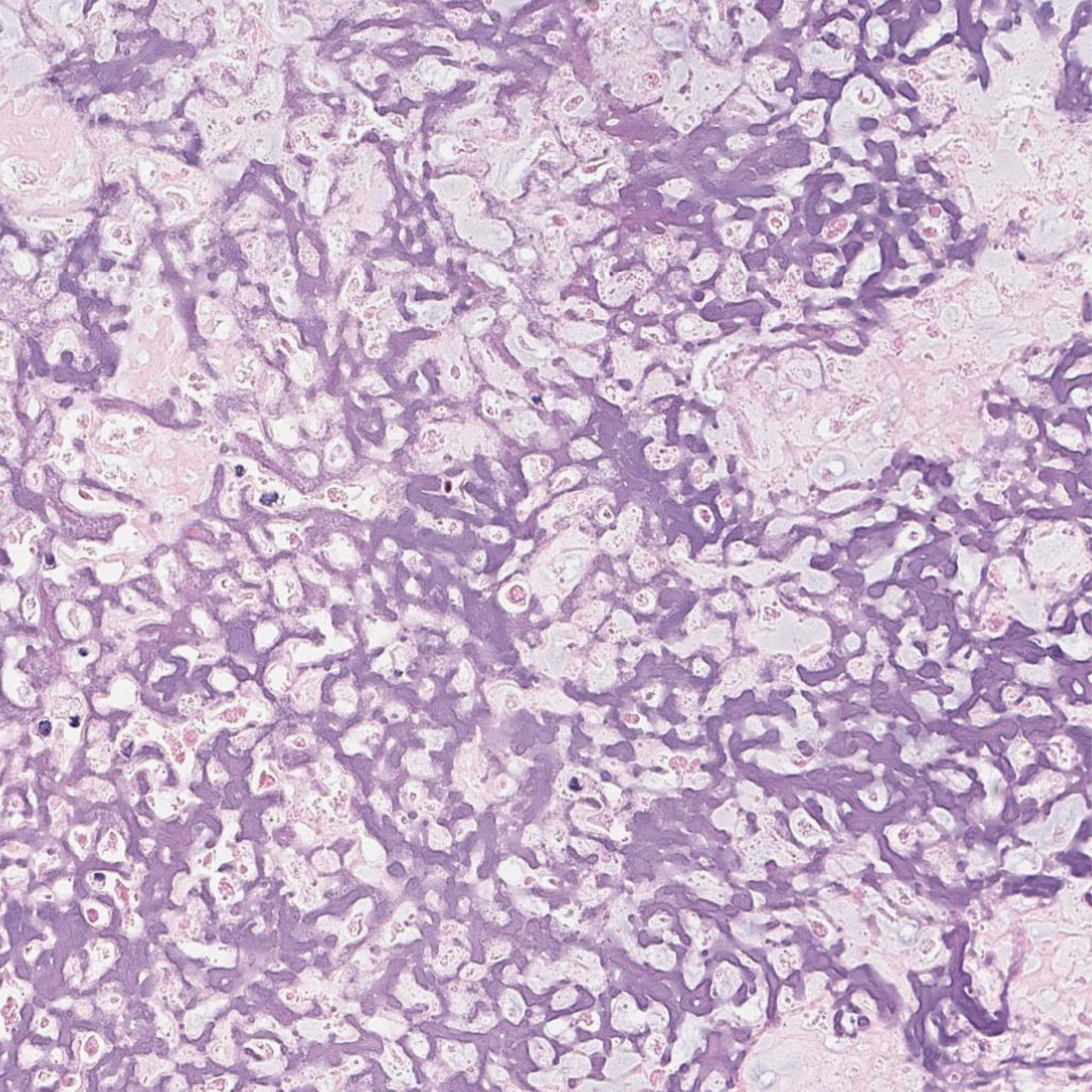,width = 0.13\textwidth}}}
\subfigure[]{\frame{\epsfig{figure=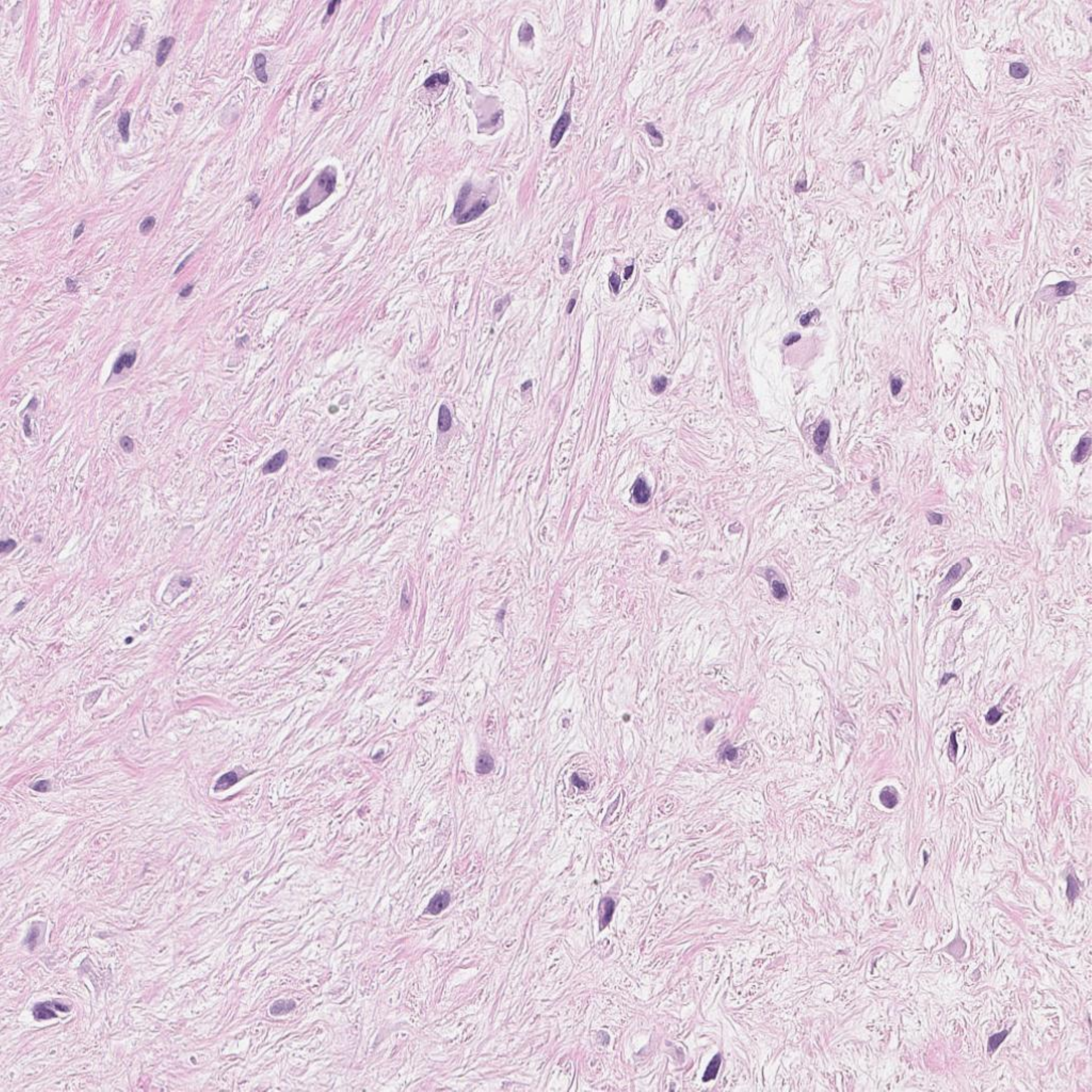,width = 0.13\textwidth}}}
\subfigure[]{\frame{\epsfig{figure=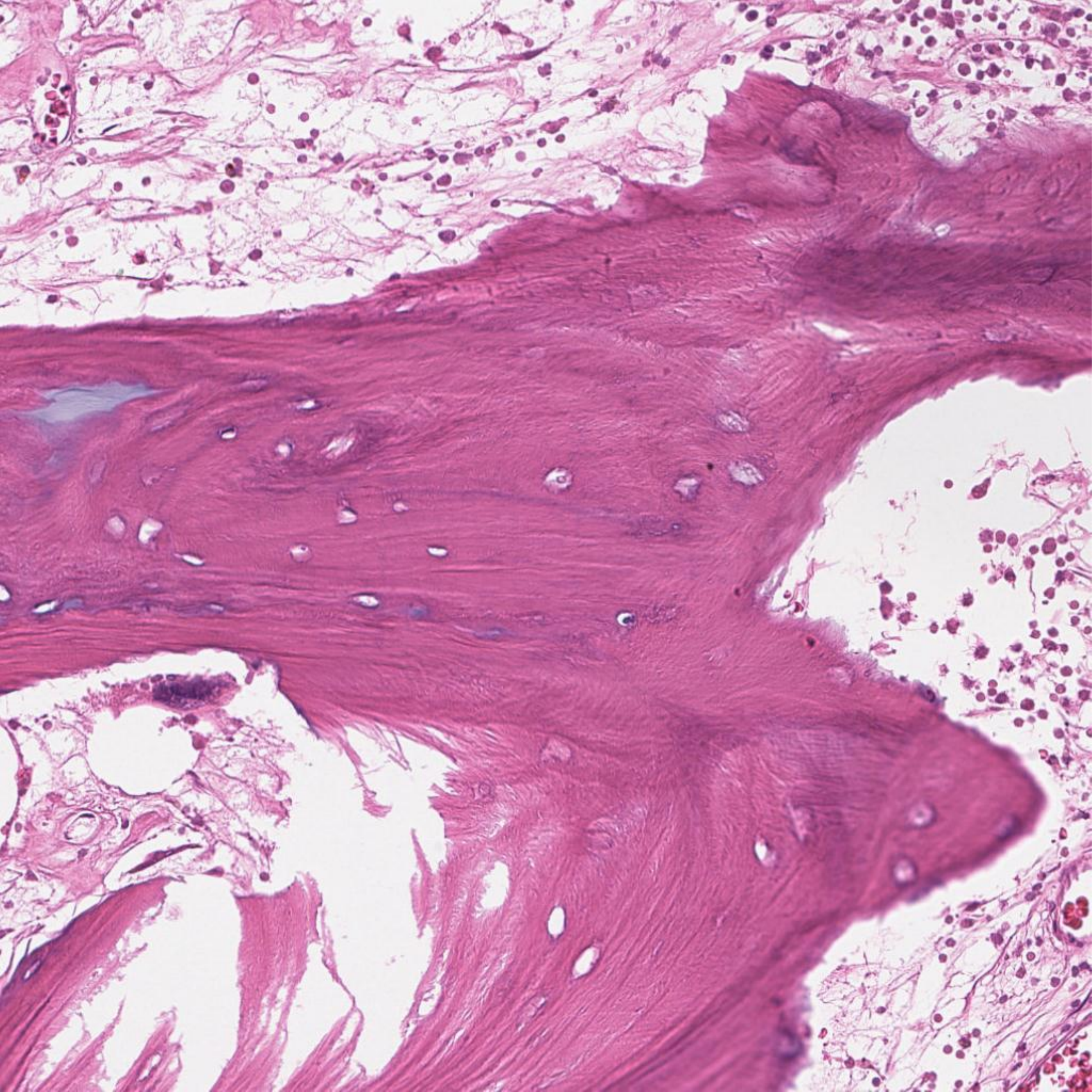,width = 0.13\textwidth}}}
\subfigure[]{\frame{\epsfig{figure=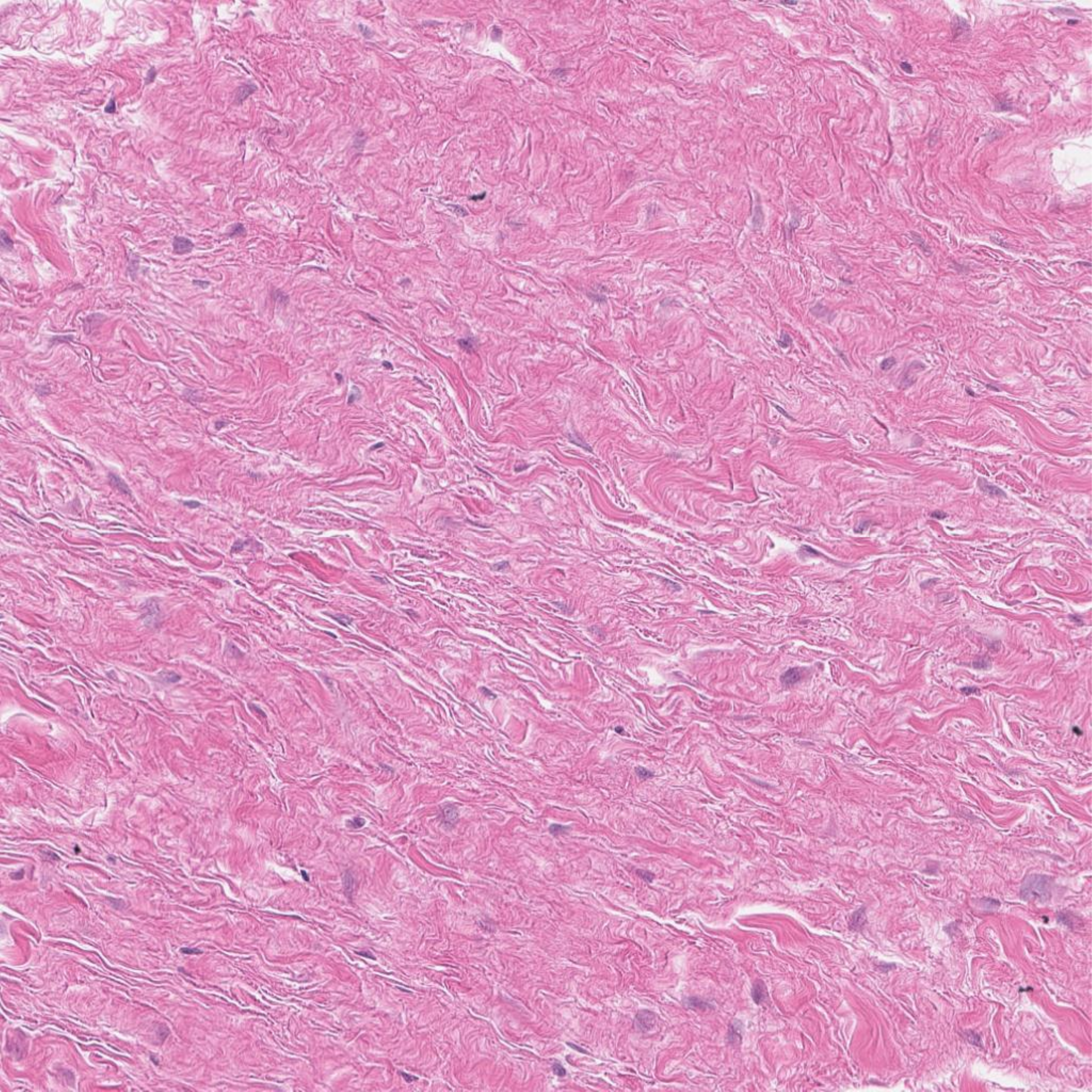,width = 0.13\textwidth}}}
\subfigure[]{\frame{\epsfig{figure=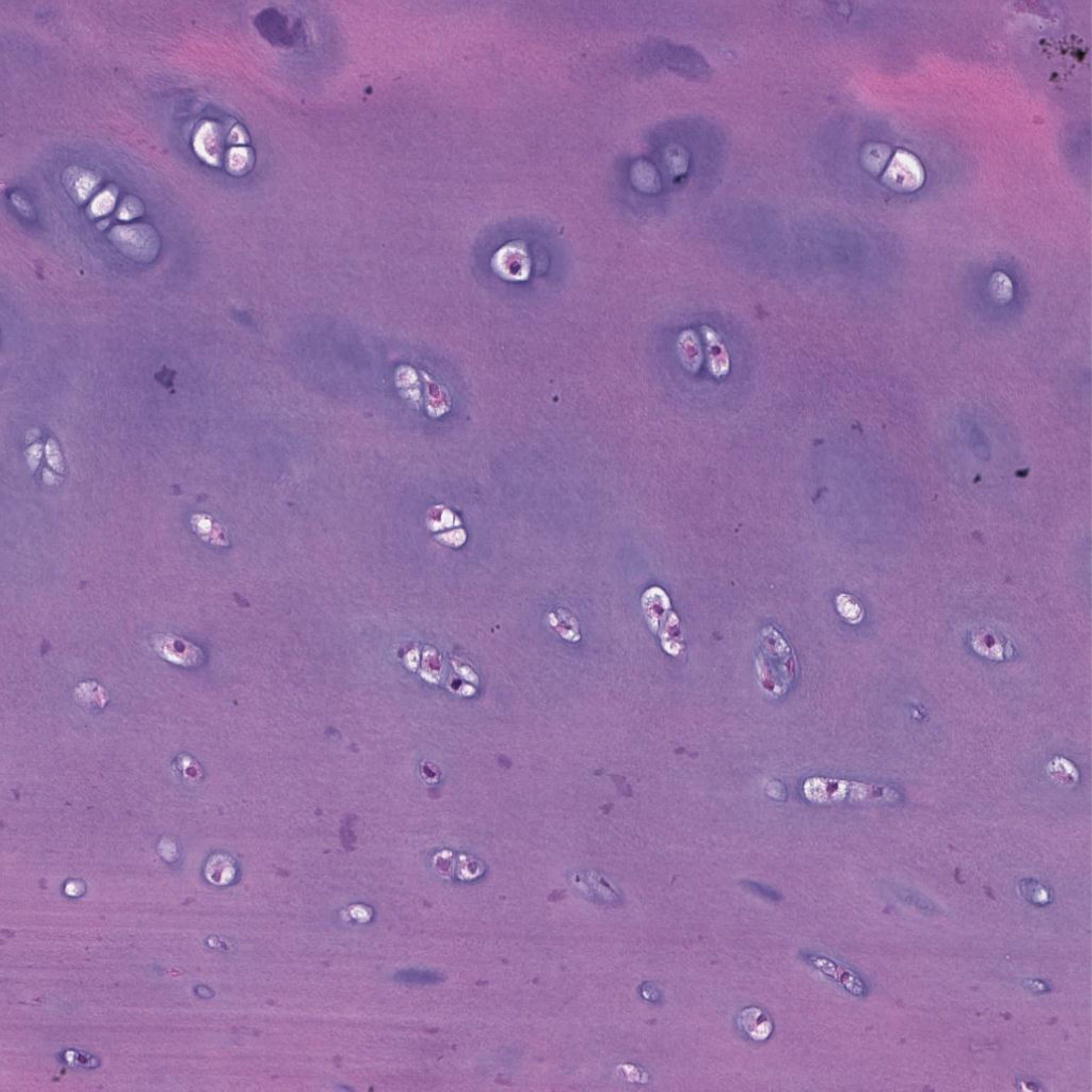,width = 0.13\textwidth}}}
\subfigure[]{\frame{\epsfig{figure=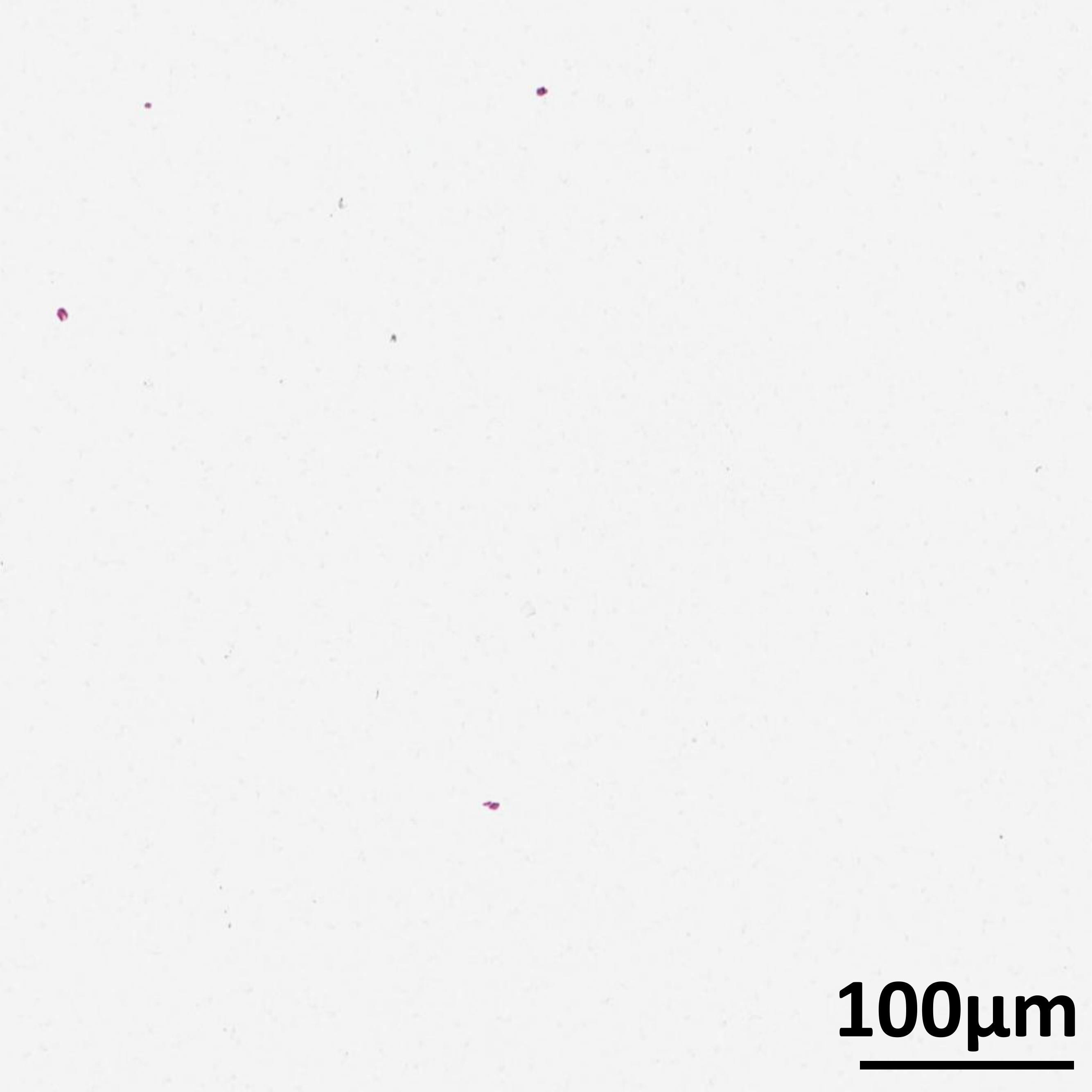,width = 0.13\textwidth}}}
\caption{A convolutional neural network we designed in this paper can predict 7 classes: (a) viable tumor, (b) necrosis with bone, (c) necrosis without bone, (d) normal bone, (e) normal tissue, (f) cartilage, and (g) blank. Our goal is to accurately segment viable tumor and necrotic tumor on osteosarcoma whole slide images for automated treatment response assessment.}
\label{fig:classes}
\end{figure}
Initial annotation on characteristic features of each class is done to train an initial CNN model.
In this work, annotators label 7 morphologically distinct classes, shown in Figure \ref{fig:classes}: viable tumor, necrosis with bone, necrosis without bone, normal bone, normal tissue, cartilage, and blank.
Note initial annotation is partially done on training images.

\subsection{Deep Interactive Learning}
During initial annotation, challenging or rare features may not be included in the training set which can lead to mislabeled predictions.
These challenging features can be added into the training set through Deep Interactive Learning (DIaL) by repeating training/finetuning, segmentation, and correction.
These three steps are repeated until annotators are satisfied with segmentation predictions on training images.

\subsubsection{Initial Training}
We need an initially trained model to annotate mislabeled regions with challenging features. WSIs are too large to be processed at once.
Thus, the labeled regions are extracted into  $256 \times 256$ pixels patches only when more than 1\% of pixels in the patch are annotated.
To balance the number of pixels between classes, patches containing rare classes are deformed to produce additional patches by elastic deformation \cite{ronneberger2015,fu2017}.
Here, we define a class is rare if the number of pixels in the class is less than 70\% of the maximum number of pixels among classes.
After patch extraction and deformation are done, some cases are separated for validating the CNN model where approximately 20\% of pixels in each class are separated.
We use a Deep Multi-Magnification Network (DMMN) \cite{ho2019} for multi-class tissue segmentation where the model looks at patches in multiple magnifications for accurate predictions.
Specifically, the DMMN is composed of three half-channeled U-Nets, U-Net-$20\times$, U-Net-$10\times$, and U-Net-$5\times$, where input patches of these U-Nets are in $20\times$, $10\times$, and $5\times$ magnifications, respectively, with size of $256 \times 256$ pixels centered at the same location.
Intermediate feature maps in decoders of U-Net-$10\times$ and U-Net-$5\times$ are center-cropped and concatenated to a decoder of U-Net-$20\times$ to enrich feature maps.
The final prediction patch of the DMMN is generated in size of $256 \times 256$ pixels in $20\times$ magnification.
To train our model initialized by \cite{glorot2010}, we use weighted cross entropy as our loss function where a weight for class $c$, $w_c$, is defined as $w_c = 1 - \frac{p_c}{\sum_{c=1}^C p_c}$, where $C=7$ is the total number of classes and $p_c$ is the number of pixels in class $c$.
Note that unlabeled regions do not contribute to the training process.
During training, random rotation, vertical and horizontal flip, and color jittering are used as data augmentation.
Stochastic gradient descent (SGD) optimizer with a learning rate of $5 \times 10^{-5}$, a momentum of 0.99, and a weight decay of $10^{-4}$ is used for 30 epochs.
In each epoch, a model is validated by mean Intersection-Over-Union (mIOU) and the model with the highest mIOU is selected as an output model.

\subsubsection{Segmentation}
After training a model is done, all training WSIs are processed to evaluate unlabeled regions.
A set of patches with size of $256 \times 256$ pixels in $20\times$, $10\times$, and $5\times$ magnifications centered at the same location is processed using the DMMN.
Note that zero-padding is done on the boundary of WSIs.
Patch-wise segmentation is repeated in $x$ and $y$-directions with a stride of 256 pixels until the entire WSI is processed.

\subsubsection{Correction}
Characteristic features are annotated during initial annotation, but challenging or rare features may not be included.
During the correction step, these challenging features that the model could not predict correctly are annotated to be included in the training set to improve the model.
In this step, the annotators look at segmentation predictions and correct any mislabeled regions.
If the predictions are satisfied throughout training images, the model is finalized.

\subsubsection{Finetuning}
Assuming the previous CNN model has already learned most features of classes, we finetune the previous model to improve segmentation performance.
Corrected regions are extracted into patches and included in the training set to improve the CNN model.
Additional patches are generated by deforming the extracted patches to give a higher weight on challenging or rare features to emphasize these features to be learned during finetuning.
SGD optimizer and weighted cross entropy with the updated weights are used during training, and we reduced a learning rate to be $5 \times 10^{-6}$ and the number of epochs to be 10 not to perturb parameters of the CNN model too much from the previous model.
Note validation cases can be selected again to utilize the majority of corrected cases for the optimization.

\subsection{Treatment Response Assessment}
The final CNN model segments viable tumor and necrotic tumor on testing WSIs.
Note necrotic tumor is a combination of necrosis with bone and necrosis without bone.
The ratio of necrotic tumor to overall tumor in case-level estimated by a deep learning model, $R^{DL}$, is defined as
\begin{equation}
    R^{DL} = \frac{p_{NT}}{p_{VT} + p_{NT}}
\end{equation}
where $p_{VT}$ and $p_{NT}$ are the number of pixels of viable tumor and necrotic tumor in a case, respectively.

\section{Experimental Results}
\begin{figure}[t!]
\centering
\subfigure[]{\frame{\epsfig{figure=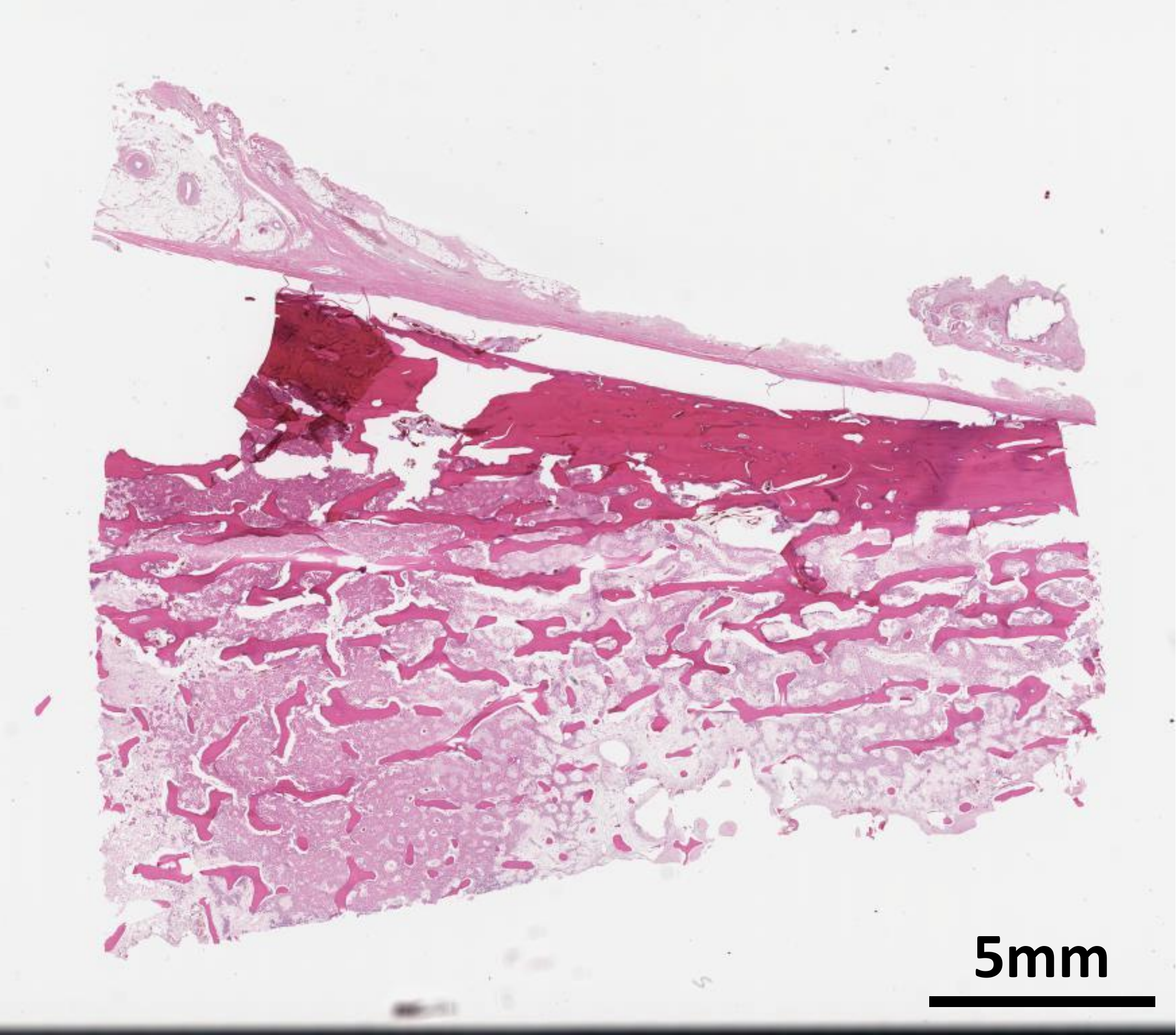,width = 0.32\textwidth}}}
\subfigure[]{\frame{\epsfig{figure=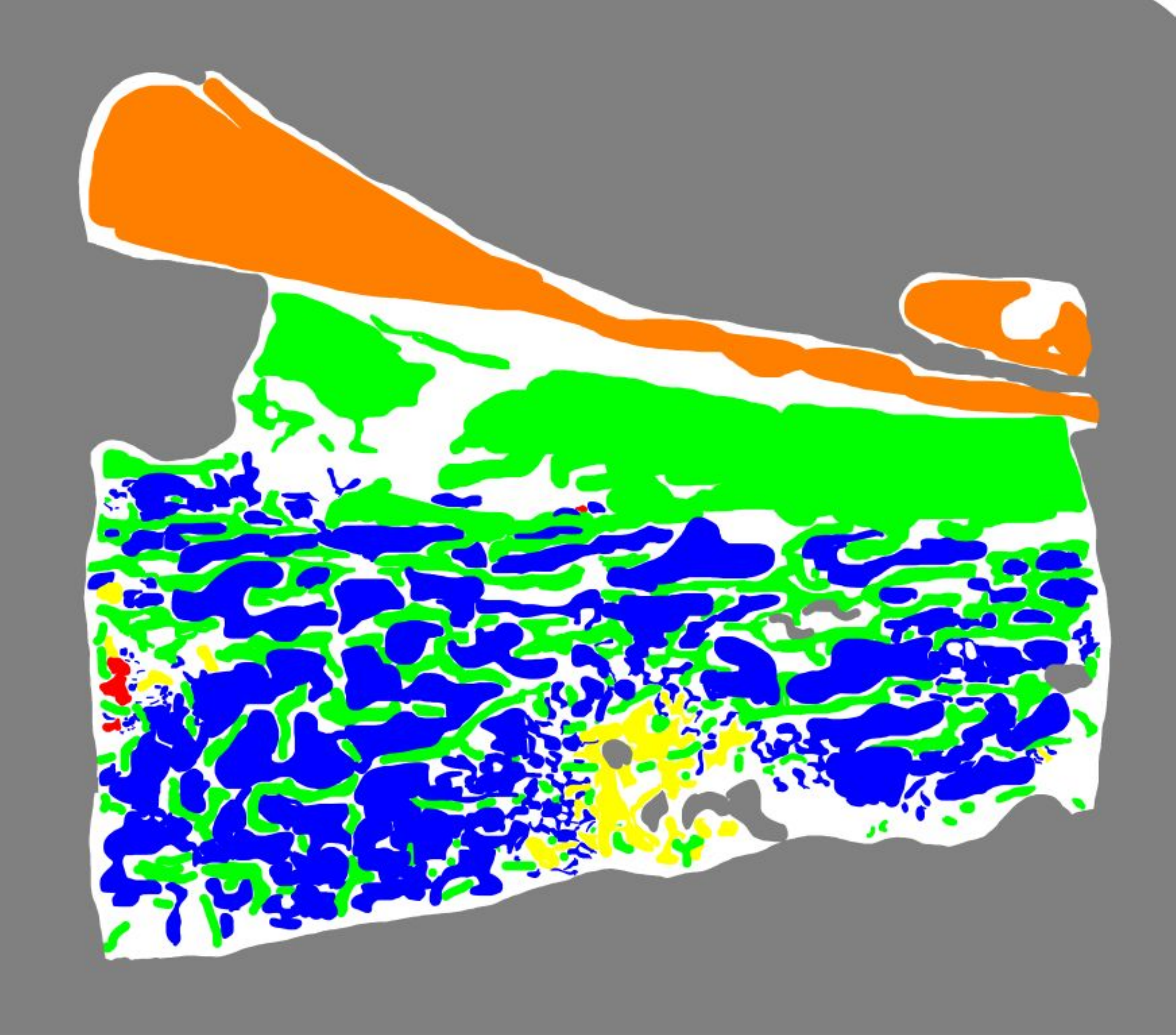,width = 0.32\textwidth}}}
\subfigure[]{\frame{\epsfig{figure=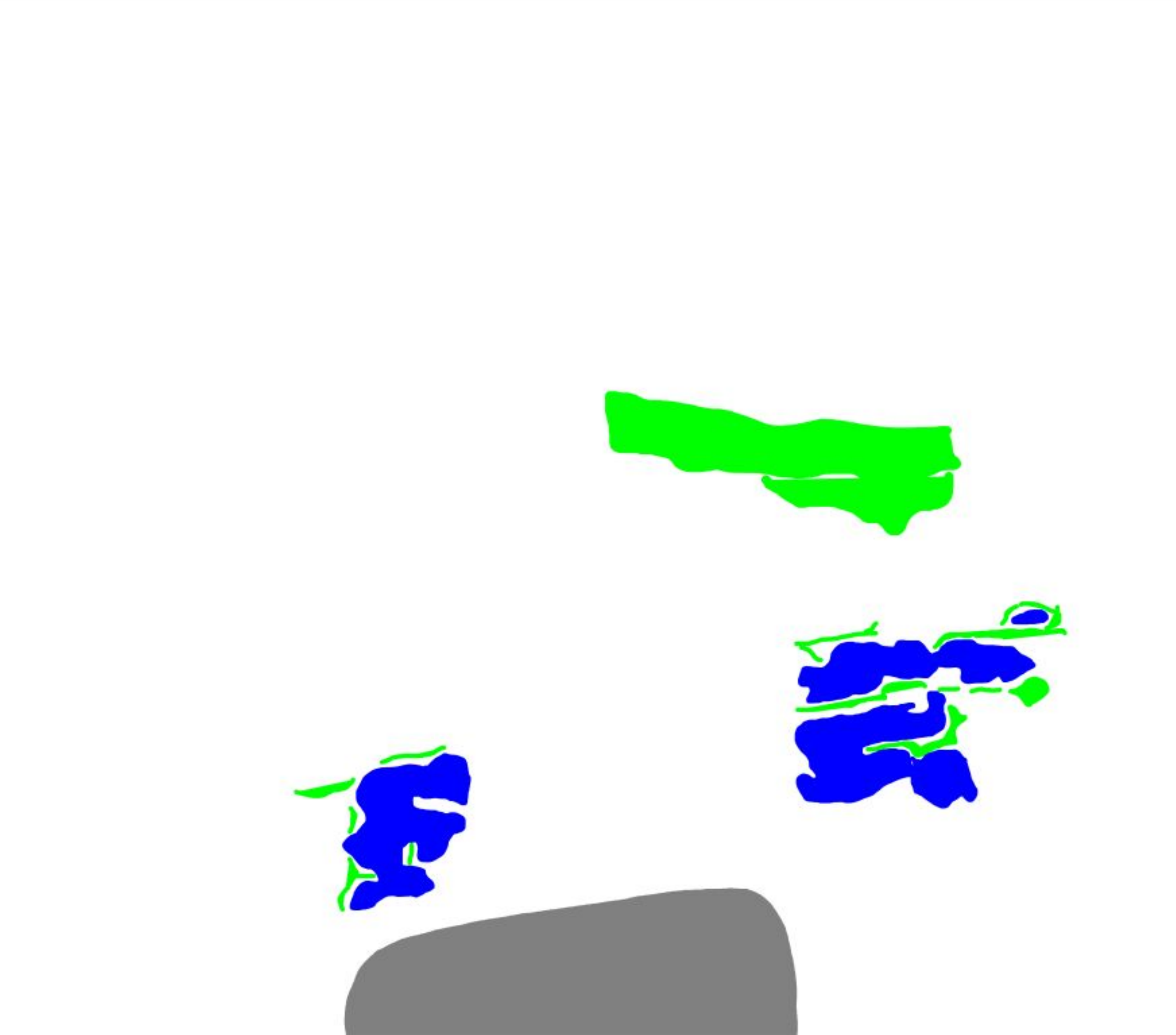,width = 0.32\textwidth}}}
\subfigure[]{\frame{\epsfig{figure=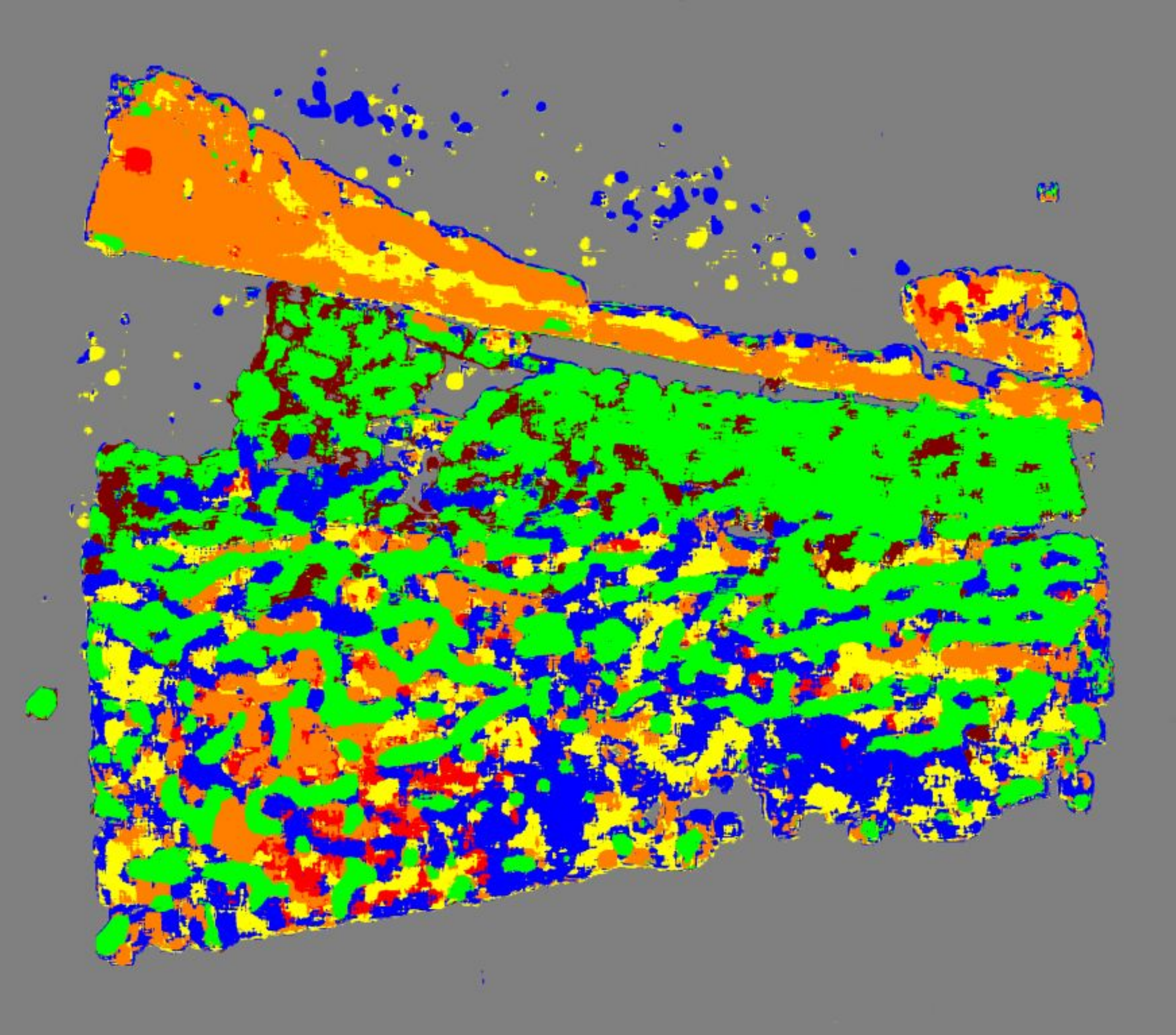,width = 0.32\textwidth}}}
\subfigure[]{\frame{\epsfig{figure=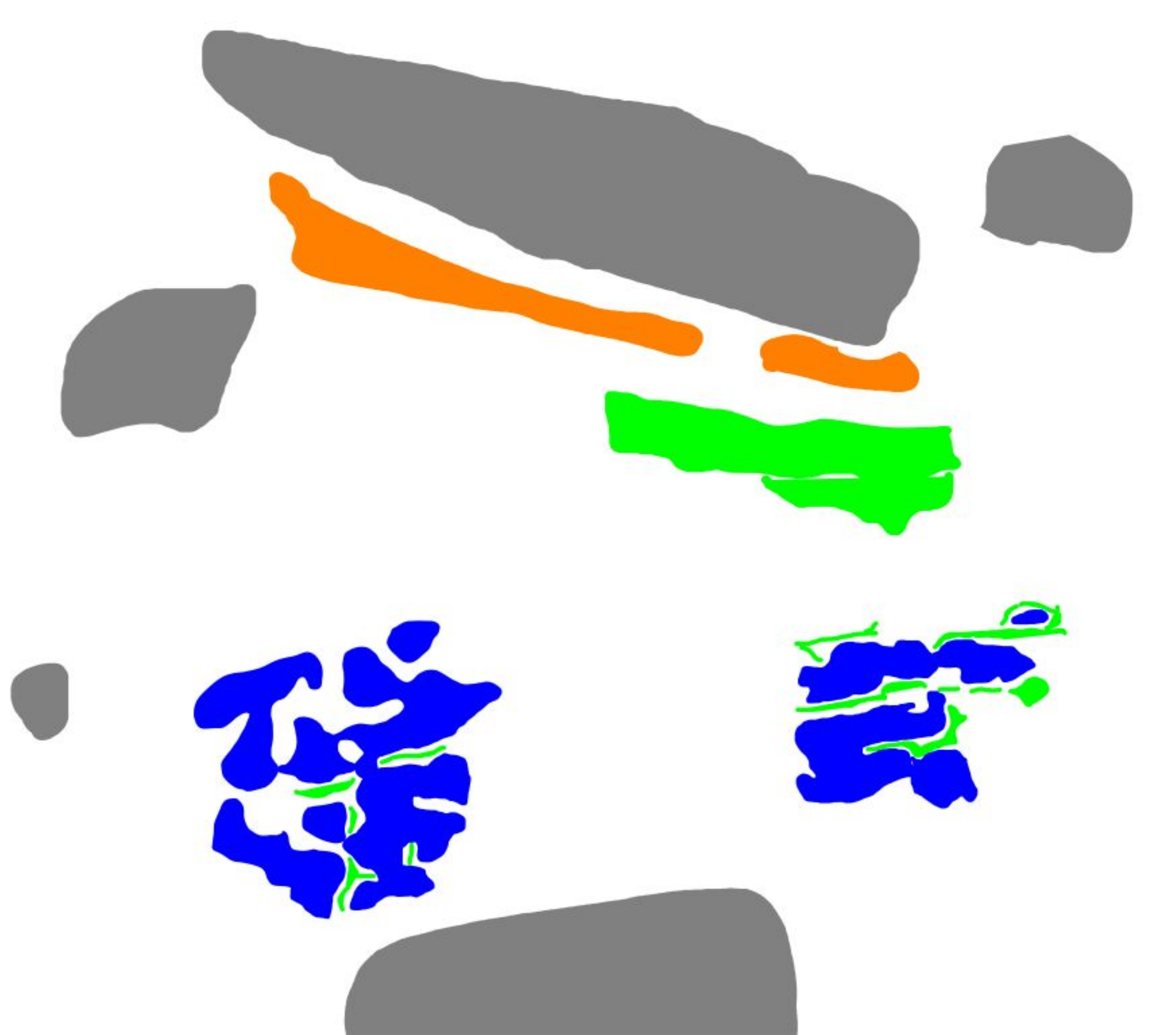,width = 0.32\textwidth}}}
\subfigure[]{\frame{\epsfig{figure=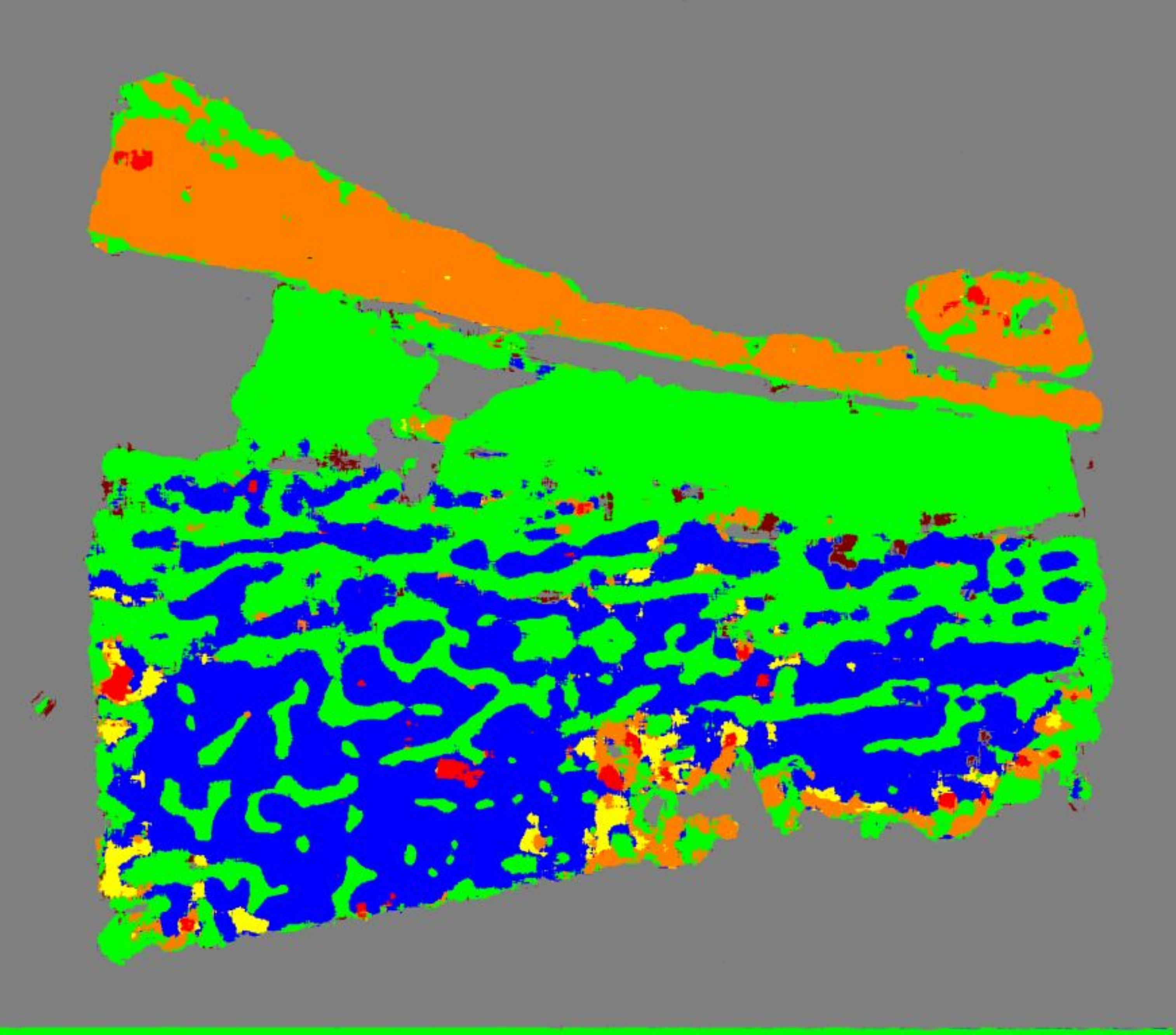,width = 0.32\textwidth}}}
\caption{An example of Deep Interactive Learning (DIaL). (a) An original training whole slide image, (b) an exhaustive annotation, (c) an initial annotation, (d) the first prediction from a CNN trained by the initial annotation, (e) the first correction where more regions for necrosis with bone, normal tissue, and blank are labeled to correct the first prediction, (f) the second prediction from a CNN finetuned from the initial model with double-weighted first correction, satisfied by annotators. The annotators spent approximately 1.5 hours to exhaustively label one whole slide image. With DIaL, the annotators are able to efficiently label characteristic features and challenging features on more diverse cases at the same given time. In this experiment, two annotators initially annotated 49 images in 4 hours and corrected 37 images in 3 hours. Note viable tumor, necrosis with bone, necrosis without bone, normal bone, normal tissue, cartilage, and blank are labeled in red, blue, yellow, green, orange, brown, and gray, respectively. White regions in (b), (c), and (e) are unlabeled regions.}
\label{fig:dial2}
\end{figure}

\begin{figure}[t!]
\centering
\subfigure[]{\epsfig{figure=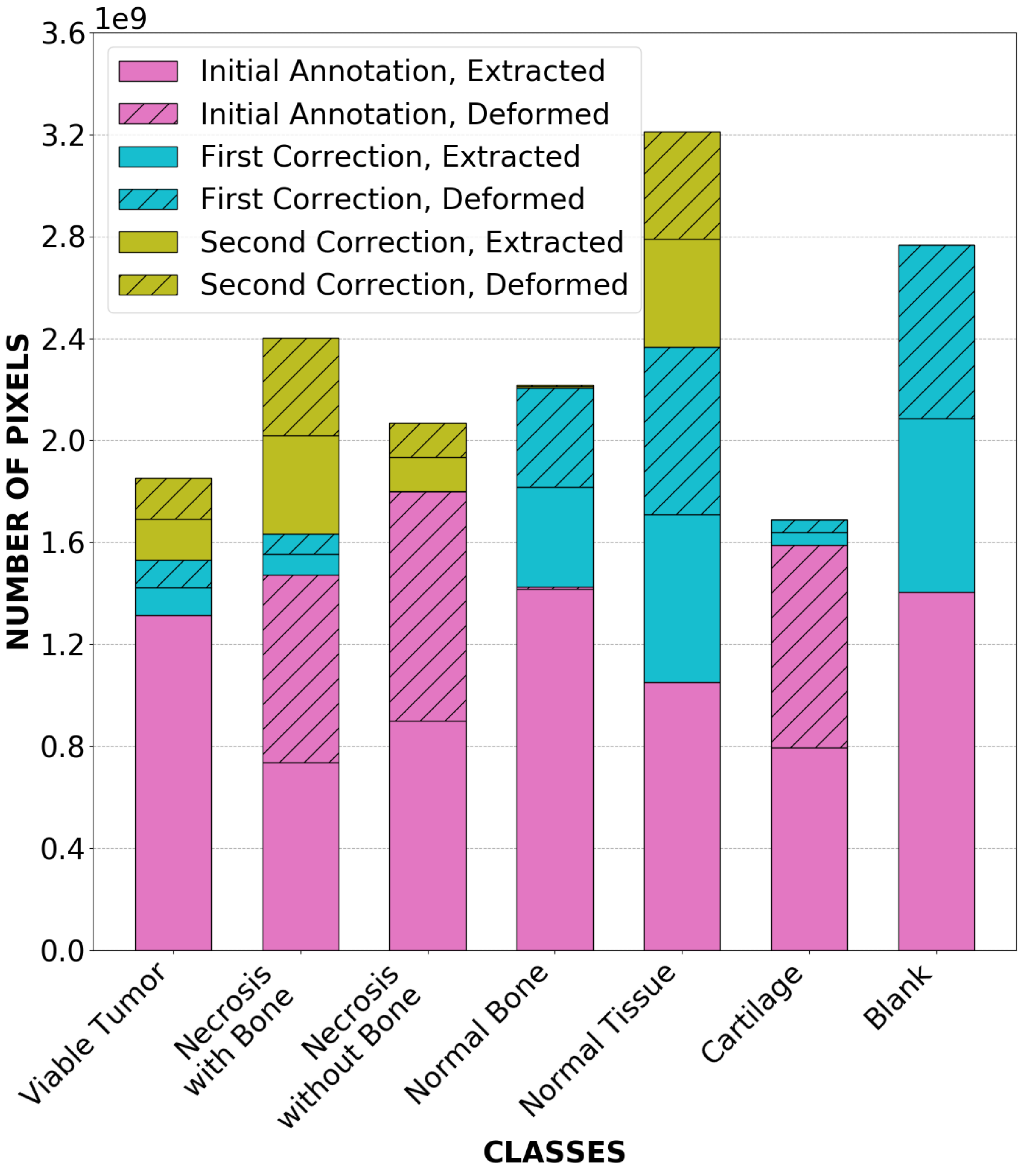,width=0.57\textwidth}}
\subfigure[]{\epsfig{figure=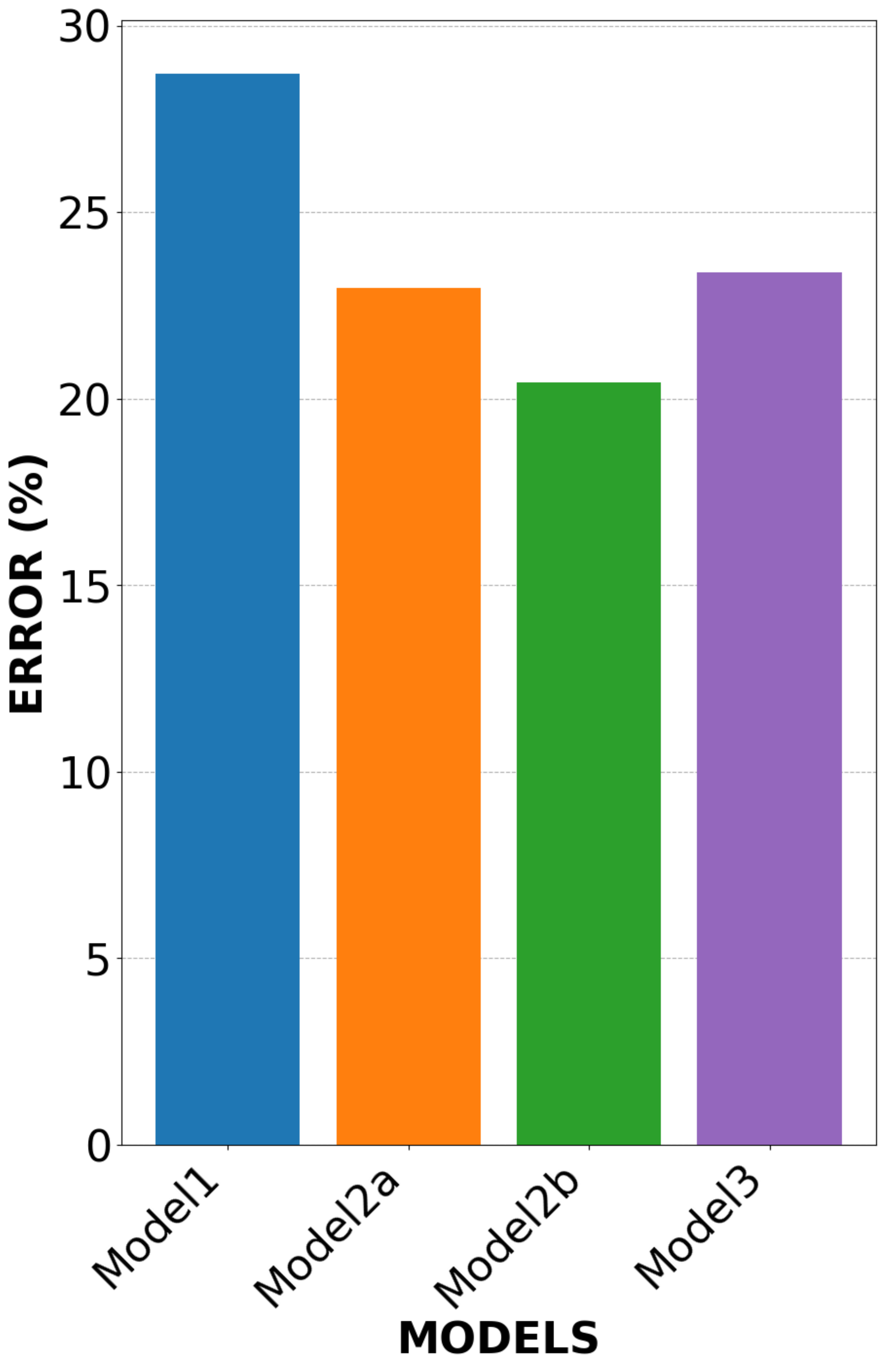,width=0.42\textwidth}}
\caption{(a) The number of pixels in a training set for each class. During initial annotation, elastic deformation \cite{ronneberger2015,fu2017} is used on patches containing necrosis with bone, necrosis without bone, and cartilage to balance the number of pixels between classes. Elastic deformation is used on all correction patches to give a higher weight on them. (b) Error rates of Model1, trained by initial annotation alone, Model2a, finetuned from Model1 with single-weighted first correction, Model2b, finetuned from Model1 with double-weighted first correction, and Model3, finetuned from Model2b with double-weighted second correction. Our final model, Model2b, achieves the error rate of 20\% considered as an expected inter-observer variation rate \cite{viray2013}.}
\label{fig:plot}
\end{figure}

Our hematoxylin and eosin (H\&E) stained osteosarcoma dataset is digitized in $20\times$ magnification by two Aperio AT2 scanners at Memorial Sloan Kettering Cancer Center where microns per pixel (MPP) for one scanner is 0.5025 and MPP for the other scanner is 0.5031.
The osteosarcoma dataset contains 55 cases with 1578 whole slide images (WSIs) where the number of WSIs per case ranges between 1 to 109 with mean of 28.7 and median of 22, and the average width and height of the WSIs are 61022 pixels and 41518 pixels, respectively.
We used 13 cases for training and the other 42 cases for testing.
Note 8 testing cases do not contain the necrosis ratio on their pathology reports, so they were excluded for evaluation.
Two annotators (N.P.A. and M.R.H.) selected 49 WSIs from 13 training cases and independently annotated them without case-level overlaps.
The pixel-wise annotation was performed on an in-house WSI viewer, allowing measuring the time taken for annotation.
The annotators labeled three iterations using Deep Interactive Learning (DIaL): initial annotation, first correction, and second correction.
They annotated 49 WSIs in 4 hours, 37 WSIs in 3 hours, and 13 WSIs in 1 hour during the initial annotation, the first correction, and the second correction, respectively.
The annotators also exhaustively labeled the entire WSI which took approximately 1.5 hours.
An example of exhaustive annotation and annotation with DIaL is shown in Figure \ref{fig:dial2}. 
With the same given time, the annotators would be able to exhaustively annotate only 5 WSIs without DIaL.
The annotators can annotate more diverse cases with DIaL.
The number of pixels annotated and deformed are shown in Figure \ref{fig:plot}(a).
The implementation was done using PyTorch \cite{paszke2019} and an Nvidia Tesla V100 GPU is used for training and segmentation.
Initial training and finetuning took approximately 5 days and 2 days, respectively.
Segmentation of one WSI took approximately $20\sim25$ minutes.

For evaluating our segmentation model, 1044 WSIs from 34 cases were segmented to estimate the necrosis ratio.
Note all WSIs were segmented as if pathologists look at all glass slides under the microscope to assess the necrosis ratio.
To numerically evaluate the estimated necrosis ratio, we compared with the ratio from pathology reports written by experts.
Here, the error rate, $E$, is defined as
\begin{equation}
E = \frac{1}{N}\sum_{i=1}^N |R_i^{PATH} - R_i^{DL}|
\end{equation}
where $R_i^{PATH}$ is the ratio from a pathology report and $R_i^{DL}$ is the ratio estimated by a deep learning model for the $i$-th case, and $1 \leq i \leq N$ where $N=34$ is the number of testing cases.
Figure \ref{fig:plot}(b) shows the error rates for our models.
Model1, Model2a, Model2b, Model3 denote an initially-trained model, a finetuned model from Model1 with single-weighted first correction, a finetuned model from Model1 with double-weighted first correction, and a finetuned model from Model2b with double-weighted second correction, respectively.
Note we tried both single-weighted correction including only extracted correction patches and double-weighted correction including both extracted correction patches and their corresponding deformed patches during the finetuning step.
We observed that the error rate decreases after the first correction, especially with a higher weight on correction patches to emphasize challenging features. 
We selected Model2b as our final model because the error rate stopped reducing after the second correction.
Our final model, trained by only 7 hours of annotations done by DIaL, was able to achieve the error rate of 20\%.
A 20\% inter-observer error rate is generally acceptable for non-standardized tasks in surgical pathology such as assessment of percentage of tumor cells has been overestimated by pathologists up to 20\% in certain instances \cite{viray2013}. 
While this cannot be directly transferred to necrosis estimation we have used this data to show that the model is able to achieve this error rate.

The task of manual quantification of the necrosis ratio done by pathologists is challenging because one must make an estimate across multiple glass slides that may differ substantially in the ratio of necrosis. 
We are convinced that our objective and reproducible deep learning model estimating the necrosis ratio within expected inter-observer variation rate can be superior to manual interpretation.

\section{Conclusion}
We presented Deep Interactive Learning (DIaL) for an efficient annotation to train a segmentation CNN.
With 7 hours of labeling, we achieved a CNN segmenting viable tumor and necrotic tumor on osteosarcoma whole slide images.
Our experiments showed that the CNN model can successfully estimate the necrosis ratio known as a prognostic factor for patients' survival for osteosarcoma in an objective and reproducible way.
In the future, we plan for patient stratification based on patients' survival data using our deep learning model.

\section{Acknowledgments/Disclosures}
This work was supported by the Warren Alpert Foundation Center for Digital and Computational Pathology at Memorial Sloan Kettering Cancer Center and the NIH/NCI Cancer Center Support Grant P30 CA008748.
T.J.F. is the Chief Scientific Officer, co-founder and equity holder of Paige.AI.
P.J.S. is a lead machine learning scientist, co-founder and equity holder of Paige.AI.
C.M.V. is a consultant for Paige.AI.
D.J.H. and T.J.F. have intellectual property interests relevant to the work that is the subject of this paper. 
MSK has financial interests in Paige.AI. and intellectual property interests relevant to the work that is the subject of this paper.
%
%
%
\bibliographystyle{splncs04}
\bibliography{mybibliography}

\end{document}